\documentclass{article}

\usepackage[T1]{fontenc}
\usepackage[utf8]{inputenc}
\usepackage{indentfirst}
\usepackage{fancyhdr}
\usepackage{graphicx}
\usepackage{lastpage}
\usepackage{float}
\usepackage{amsmath}
\usepackage{setspace}
\usepackage{enumitem}
\usepackage{mathpazo}
\usepackage{booktabs}
\usepackage{tabto}
\usepackage{xcolor}
\usepackage{multirow}
\usepackage{microtype}
\usepackage{upgreek}
\usepackage{amsthm}
\usepackage{hyphenat}
\usepackage[sort&compress]{natbib}
\usepackage[hidelinks]{hyperref}
\usepackage{cleveref}
\usepackage{url}
\usepackage{geometry}
\usepackage{newfloat}
\usepackage{caption}

\usepackage{bm}

\newcommand{\beq}{\begin{equation}}
\newcommand{\eeq}{\end{equation}}
\newcommand{\aligneqn}[1]{\begin{align}#1\end{align}}

\renewcommand{\vec}[1]{\mathbf{#1}}
\newcommand{\unitv}[1]{\hat{\bf #1}}

\newcommand{\paren}[1]{\left( #1 \right)}
\newcommand{\bracket}[1]{\left[ #1 \right]}
\newcommand{\braces}[1]{\left\{ #1 \right\}}
\newcommand{\vr}{\vec{r}}
\newcommand{\vp}{\vec{p}}
\newcommand{\vv}{\vec{v}}
\newcommand{\vk}{\vec{k}}
\newcommand{\veta}{\bm{\upeta}}
\newcommand{\twov}[2]{\left[\begin{array}{c} #1 \\ #2\end{array}\right]}

\newcommand{\changeurlcolor}[1]{\hypersetup{urlcolor=#1}}   

\begin{document}

\title{Semiclassical Phase Analysis for a Trapped-Atom Sagnac Interferometer}

\author{Zhe Luo,  E R Moan, and C A Sackett \\
Physics Department, University of Virginia, Charlottesville, VA 22904}

\maketitle

\begin{abstract}
{A Sagnac atom interferometer can be constructed using a Bose-Einstein condensate trapped in a cylindrically symmetric harmonic potential.
Using the Bragg interaction with a set of laser beams, the atoms can be launched into circular orbits, with two counterpropagating interferometers 
allowing many sources of common-mode noise to be excluded. In a perfectly symmetric and harmonic potential, the interferometer output would depend
only on the rotation rate of the apparatus. However, deviations from the ideal case can lead to spurious phase shifts. These phase shifts
have been theoretically analyzed for anharmonic perturbations up to quartic in the confining potential, as well as angular deviations of the laser beams,
timing deviations of the laser pulses, and motional excitations of the initial condensate. 
Analytical and numerical results 
show the leading effects of the perturbations to be second order. The scaling of the phase shifts with the number of orbits and
the trap axial frequency ratio are determined. The results indicate that sensitive parameters should be controlled at the $10^{-5}$ level to accommodate a
rotation sensing accuracy of $10^{-9}$ rad/s. The leading-order perturbations are suppressed in the case of perfect cylindrical symmetry, even in the presence
of anharmonicity and other errors. An experimental measurement of one of the perturbation terms is presented.}
\end{abstract}


\section{Introduction}

Atom interferometry is useful for many types of precision measurements \cite{Berman1997,Cronin2009,Tino2014}, 
but one of the most attractive applications is
inertial navigation \cite{Grewal2013,Geiger2020}. For this purpose, atom interferometers can be configured to measure 
accelerations \cite{McGuirk2002,Schmidt2011}, rotations \cite{Durfee2006,Savoie2018}, or both 
simultaneously \cite{Dickerson2013,Chen2019}. 
In order to obtain high sensitivity, it is desirable to operate the interferometer with long
measurement times $T$. This can be a challenge for measurements with freely falling atoms, since a large fall distance
will increase the apparatus size and complexity. One solution is to support the atoms against gravity using a magnetic
field \cite{GarridoAlzar2019}. This unavoidably leads to some confinement as well \cite{Sackett2006}, but we can make use of this confining potential
to help guide the atoms along a desired trajectory. 

In Ref.~\cite{Moan2020}, we demonstrated a Sagnac interferometer using atoms confined in a harmonic potential, where the trap caused
the atoms to move in nearly circular orbits so as to enclose an area. By using two simultaneous counter-propagating
interferometers in the same trap, many spurious phase shifts can be rejected through a differential measurement. However,
if the confining potential is not ideal then it can still impact the final phase measurement and limit the accuracy of the sensor.
In this paper, we analyze the phase shifts imparted by the potential and present the dependence on various perturbative terms. 
We find that it is possible to reach performance levels consistent with precision navigation requirements, but that doing
so will require precise control of the trapping parameters. 

The analysis described here is similar to that by West in Ref.~\cite{West2019}. However, in that work only a single interferometer 
was considered. By design, many errors cancel in the dual-interferometer scheme of \cite{Moan2020}, making our approach 
mainly sensitive to higher-order effects that were not
comprehensively addressed in \cite{West2019}. In addition, our numerical analysis
considers a wider variety of perturbative parameters, including ones that couple all three dimensions. 

The analysis here focuses exclusively on the phase shifts induced by the perturbations. The interferometer visibility and enclosed
area are also impacted by non-idealities. These are not affected by the differential measurement, so the
analysis in \cite{West2019} is directly applicable. While the visibility and area are important, we expect
the phase to be the most sensitive parameter, so if perturbations are reduced to the point that phase shifts are negligible, 
the visibility and area will also be close to ideal. 

In the following, we first present our interferometer scheme and 
the semiclassical approach we use for the analysis. We apply the approach analytically 
to the case of a harmonic trapping potential with a limited number of perturbations. We then present numerical results
for a broader set of perturbations. Finally, we discuss the implications for experiments and compare to
a measurement of the phase sensitivity for one pair of parameters. 
 
\section{Semiclassical Phase Analysis}

\begin{figure}
\includegraphics[width=7 cm]{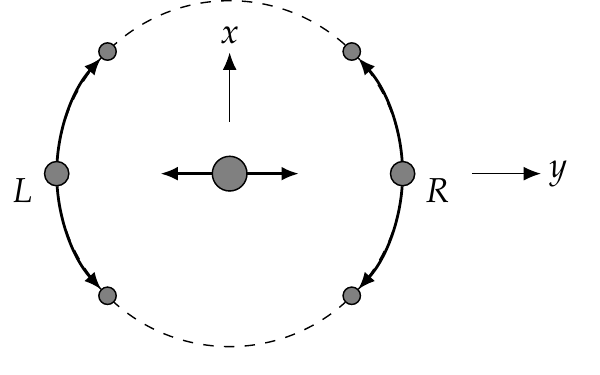}
\caption{Trajectory of atoms in Sagnac interferometer. A condensate, represented by gray disks, starts in the center of
	the harmonic trap at time $t=0$. A standing-wave laser oriented along $y$ splits the condensate into two packets that move 
	in the $\pm y$ directions. One quarter period later, the atoms come to rest on the left (L) 
	and right (R) sides of the trap.
	Another standing-wave laser oriented along $x$ then splits the atoms again.
	The resulting packets have the correct velocity to travel in a circular orbit around the trap center. After $n$ complete orbits,
	the $x$-laser is applied again. This brings some of the moving atoms back to rest on their respective sides of the trap.
	The fraction brought to rest depends on the phase difference between the two interfering packets.
	\label{figtraj}}
\end{figure}

The atom trajectories used in our interferometer scheme are illustrated in Fig.~\ref{figtraj}. A Bose condensate is first prepared nominally 
at rest in the center of an approximately harmonic trap. We define a coordinate system centered on the trap, with $z$ vertical.
An off-resonant standing wave laser is applied along the $y$ direction, driving the atoms into a superposition of states
moving at momenta $\vp = \pm 2\hbar k\unitv{y}$, where $k$ is the wave number of the laser \cite{Wu2005a}. The two wave packets separate,
and after one quarter of an oscillation period they come to rest on opposite side of the trap. At that time another standing wave
is applied along the $x$ direction, resulting in a total of four wave packets. The momentum kicks along $x$ cause each packet
to move in a nearly circular orbit around the center of the trap. The packets are sufficiently dilute that they can pass through each
other with negligible losses. After one or more orbits, the $x$ standing wave is applied again, 
closing each of the two interferometers. If the quantum state of two packets prior to the recombination pulse is
\beq
\psi \approx \frac{1}{\sqrt{2}}\bracket{e^{i\phi_+} e^{2ikx} + e^{i\phi_-}e^{-2ikx}},
\eeq
then the probability for an atom to be brought to rest by the recombination pulse is $\cos^2(\Delta\phi/2)$ for 
phase difference $\Delta \phi = \phi_+-\phi_-$; atoms not brought to rest continue moving. After 
a short time of flight, the atoms can be imaged and the fraction of the population in each momentum state can be determined. 

One contribution to the phase is the Sagnac effect, giving $\Delta\phi  = 4m\Omega A/\hbar$ 
where $A$ is the area enclosed by an orbit,
$m$ is the atomic mass, and $\Omega$ is the rotation rate of the apparatus. This phase applies with opposite signs to the
two interferometers, so if $\Delta\phi_R$ denotes the phase measured at positive $y$ and $\Delta\phi_L$ denotes the phase measured
at negative $y$, we have $\Phi \equiv \Delta\phi_R-\Delta\phi_L = 8m\Omega A/\hbar$. 
This is the signal used for rotation sensing.

Our goal in this paper is to analyze other contributions to this differential phase, so we assume $\Omega = 0$. The remaining
phase for a single interferometer is given in the semiclassical approximation by a sum of three terms \cite{Hogan2009}:
\beq \label{allphase}
\Delta\phi = \Delta\phi^{\mathrm{dyn}} + \phi^{\mathrm{laser}} + \phi^{\mathrm{sep}} .
\eeq
The first is the dynamical phase acquired by the atoms as they move through the trap. Each packet acquires a phase 
$(1/\hbar) \int L\,dt$, for Lagrangian $L = mv^2/2-V(\vr)$ evaluated on the classical trajectory of the packet. Here $V(\vr)$ is the trapping
potential. We therefore express
\beq \label{phidyn}
\Delta\phi^{\mathrm{dyn}} = \frac{1}{\hbar} \int_{t_a}^{t_b} (L_+ - L_-)\,dt,
\eeq
where the $x$-splitting pulse is applied at time $t_a$ and the recombination pulse at time $t_b$.
The $\pm$ labels refer to the wave packet that initially received a Bragg kick in the $\pm x$ direction,
with $L_\pm$ indicating the Lagrangian evaluated along that packet's trajectory.

The laser phase is set by the
location of the atoms relative to the Bragg standing wave. If the atoms are initially split at position $\vr_a$ and
recombined at position $\vr_b$, then the interferometer registers a phase shift
\beq \label{laserphase}
\phi^{\mathrm{laser}} = 2\vk\cdot(\vr_a-\vr_b).
\eeq
An additional phase shift will appear if the position of the Bragg standing wave itself changes between the two pulses,
but we omit this effect since the additional shift will be the same for both interferometers and thus cancel in $\Phi$. If the 
interferometer is not closed, such that the final positions of the packets are $\vr_{b+}$ and $\vr_{b-}$, then we evaluate
the phase at the center position $\vr_b \rightarrow (\vr_{b+}+\vr_{b-})/2$. 

The separation phase $\phi^{\mathrm{sep}}$ also
arises when the interferometer is not closed, and it accounts for the fact that if the final packets are not at rest, their phases are themselves
position-dependent. This results in 
\beq \label{sepphase}
\phi^{\mathrm{sep}} = - \frac{m}{2\hbar}(\vv_{b+}+\vv_{b-})\cdot(\vr_{b+}-\vr_{b-})
\eeq
where $\vv_{b\pm}$ are the final wave packet velocities.

For a single atom in a harmonic potential, the semiclassical approximation is very accurate since it agrees with the fully quantum
results obtained using coherent-state wave functions. However, interacting atoms in a condensate will occupy a Thomas-Fermi wave function
which can be quite different from a coherent state \cite{Dalfovo1999}. In addition, the wave function can be distorted by anharmonic perturbations. These
effects are not accounted for in the semiclassical approach. We are currently exploring this issue using an approach based on the Gross-Pitaevskii equation
for an interacting condensate \cite{Edwards2020,Stickney2021}. Preliminary results indicate that the semiclassical approximation is 
still quite accurate as long as the anharmonic perturbations are small. 
Use of a realistic wave function might, however, have more significant impact on the visibility and effective enclosed area.

\section{Harmonic Oscillator Potential}

For a harmonic trap, the classical trajectories can be expressed analytically, which allows an analytic calculation 
of the final differential phase $\Phi$. We consider a potential of the form
\beq \label{eq:potential}
V(\vr) = \frac{1}{2}m\paren{\omega_1^2 r_1^2 + \omega_2^2 r_2^2 + \omega_3^2 r_3^2}.
\eeq
Here the $r_i$ are the principal
coordinates of the trap, in which the potential has this diagonal form. The $\omega_i$ are the corresponding oscillation
frequencies. The principal coordinates 
do not in general conform to the coordinates by which the Bragg laser beams are defined as in the previous section.
We work in the principal coordinates and take the Bragg wave vectors as
\beq 
\vk_x = k \sum_i \kappa_{xi} \unitv{e}_i \qquad \qquad  \vk_y = k \sum_i \kappa_{yi} \unitv{e}_i
\eeq
for principal basis vectors $\unitv{e}_i$ and unit vector components $\kappa_i$. 

If at time $t_a$ an atom has position $\vr_a$ and velocity $\vv_a$, its subsequent trajectory is given by 
\beq
r_i(t) = r_{ai} \cos \omega_i(t-t_a) + \frac{v_{ai}}{\omega_i} \sin\omega_i(t-t_a)
\eeq
and
\beq
v_i(t) = -\omega_i r_{ai} \sin \omega_i(t-t_a) + v_{ai} \cos \omega_i(t-t_a).
\eeq
Since the Lagrangian is separable in the principal coordinates, the dynamical phase can be calculated for each
coordinate independently, and can be evaluated as 
\beq
\phi^{\mathrm{dyn}}_i(t) = \frac{m}{4\hbar\omega_i} \braces{\paren{v_{ai}^2-\omega_i^2 r_{ai}^2} \sin 2\omega_i (t-t_a)
+ 2\omega_i v_{ai} r_{ai} \big[ \cos 2\omega_i(t-t_a) - 1\big] }.
\eeq

To apply this result, we must express the trajectory in our interferometer. We suppose the initial condensate has
position $\vr_0$ and velocity $\vv_0$. The $y$ Bragg laser is applied at time zero, and the atoms are allowed to propagate for
time $t_1$. The wave packets then have coordinates
\beq
r_{si} = r_{0i} \cos \omega_i t_1 + \frac{1}{\omega_i}(v_{0i} + s v_B \kappa_{yi})\sin\omega_i t_1,
\eeq
where $v_B \equiv 2\hbar k/m$ is the velocity kick from the Bragg beam. 
We take $s = \pm 1$ for atoms kicked in the $\pm y$ directions. 
To help keep the notation clear, we label $s = +1$ with $R$ and $s=-1$ with $L$.
The starting velocities for the actual interferometers also include the $x$-Bragg kicks, leading to trajectories
\beq \label{positions}
r_{s\pm i} = r_{si} \cos \omega_i(t-t_1) + \frac{1}{\omega_i}(v_{si}\pm v_B \kappa_{xi}) \sin \omega_I(t-t_1).
\eeq
Here we use the $+$ and $-$ symbols to label the sign of the kicks received from the $x$ beams.
The atoms orbit for a time $t_2$, so the total duration of the motion is $t_1 + t_2$.

We use these trajectories to evaluate the phase. We set $\Phi^{\textrm{dyn}}_i$ as the dynamical differential phase
$\Phi_i^{\mathrm{dyn}} = (\phi^{\mathrm{dyn}}_{R+i} - \phi^{\mathrm{dyn}}_{R-i}) - (\phi^{\mathrm{dyn}}_{L+i}-\phi^{\mathrm{dyn}}_{L-i})$ and find
\beq \label{Phid}
\Phi_i^{\mathrm{dyn}} = \frac{4kv_B}{\omega_i} \kappa_{xi} \kappa_{yi} \big[\sin \omega_i (2t_2 + t_1) - \sin \omega_i t_1\big].
\eeq
The separation phase of Eq.~\eqref{sepphase} can be evaluated in terms of the final positions
and velocities. We obtain
\beq
\Phi_i^{\mathrm{sep}} = -\frac{4kv_B}{\omega_i} \kappa_{xi} \kappa_{yi} \big[\sin \omega_i (2t_2 + t_1) - \sin \omega_i t_1\big],
\eeq
so this term exactly cancels the dynamical phase here. The net differential phase is therefore given by
$\Phi^{\mathrm{laser}}$, which is evaluated to be
\beq
\Phi = - 4kv_B \sum_i \frac{\kappa_{xi} \kappa_{yi}}{\omega_i} \big[\sin \omega_i(t_1+t_2) -\sin \omega_i t_1\big]
\eeq

The ideal case consists of a cylindrically symmetric trap with $\omega_1 = \omega_2 \equiv \omega$ in the horizontal directions,
and $\omega_3 = \omega_z \equiv \zeta\omega$ 
perhaps different in the vertical direction. The Bragg beams should have $\kappa_{x1} = \kappa_{y2} = 1$,
with the other $\kappa$ components equal to zero.  We then obtain
$\Phi = 0$ and there is no phase shift from the trap potential. To allow for small
deviations from the ideal case, we consider a nearly-symmetric potential of the form
\beq \label{harmpot}
V = \frac{1}{2}m\omega^2\bracket{\paren{1+\Delta}x^2 + \paren{1-\Delta}y^2 + 2\gamma xy + \zeta^2 z^2},
\eeq
with $|\Delta|$ and $|\gamma|$  small compared to one. This can be diagonalized to give principal horizontal frequencies
\beq
\omega_{1,2} = \omega\sqrt{1\pm \Gamma} \approx \omega\paren{1\pm \frac{\Gamma}{2}},
\eeq
where $\Gamma = \sqrt{\Delta^2+\gamma^2}$. We take $\omega_1$ to use the plus sign and $\omega_2$ to use
the minus sign. The principal directions can then be expressed
\beq
\unitv{e}_1 = \frac{1}{\sqrt{2\Gamma(\Gamma-\Delta)}} \twov{\gamma}{\Gamma-\Delta} 
\qquad \text{and} \qquad
\unitv{e}_2 = \frac{1}{\sqrt{2\Gamma(\Gamma+\Delta)}} \twov{-\gamma}{\Gamma+\Delta} .
\eeq

We also allow the Bragg beams to deviate slightly from
their nominal alignments, with 
\aligneqn{
\vk_x & = k\paren{\unitv{x} + \psi_x' \unitv{y} + \psi_x''\unitv{z}} \label{kx} \\
\vk_y &  = k\paren{-\psi_y'\unitv{x} + \unitv{y} + \psi_y''\unitv{z}} \label{ky}
}
for $|\psi_j'|, |\psi_j''| \ll 1$. Finally, we allow for timing errors, defining
\beq
\delta_1 = \omega t_1 - \frac{\pi}{2} \qquad \text{and} \qquad \delta_2 = \omega t_2 - 2\pi n
\eeq
for an interferometer with $n$ orbits.

We then expand the total phase $\Phi$ to second order in the small parameters. We obtain
\beq \label{phi_appx}
\Phi \approx \frac{4\pi k v_B}{\omega} \gamma \bracket{ n\delta_1 + \paren{n+\frac{1}{4}}\delta_2}
- \frac{4kv_B}{\zeta\omega} \psi_x'' \psi_y'' \bracket{\sin2\pi\zeta \paren{n+\frac{1}{4}} - \sin \frac{\pi \zeta}{2}}
\eeq
Note that the radius of the orbit which the atoms undergo is  $R = v_B/\omega$, so the prefactors scale as  $kR$,
which is very large
for orbits of mm or cm size. We see that the critical parameters are the timing, the $xy$ term in the potential,
and the vertical alignment of the Bragg beams. There is no first-order dependence on small parameters.

\section{Anharmonic potential}

More generally, the trapping potential will not be perfectly harmonic, so it is important to understand the impact of
small anharmonic terms. Anharmonic perturbations make analytical calculations complicated \cite{Landau1976}, so here we use
a numerical approach. We consider a trapping potential of the form
\beq \label{potentialterms}
V(x,y,z) = \frac{1}{2}m\omega^2\paren{x^2+y^2+\zeta^2 z^2} 
+ \frac{1}{2}mv_B^2 \sum_{\lambda \mu \nu}  \frac{c_{\lambda\mu\nu} x^\lambda y^\mu z^\nu}{R^{\lambda+\mu+\nu}},
\eeq
with the $c_{\lambda\mu\nu}$ coefficients dimensionless and small compared to 1. We include corrections from second to fourth order,
$2\leq \lambda + \mu + \nu \leq 4$, noting that first-order terms can always be eliminated by offsetting the location of the trap minimum.

Other perturbative parameters are the initial position $\vec{r}_0/R$, the initial velocity $\vv_0/v_B$,
the Bragg alignment errors $\psi_i', \psi_i''$ from Eqs.~\eqref{kx} and \eqref{ky}, and the timing errors $\delta_1$, $\delta_2$. 
Because the perturbations can change the effective oscillation frequency, we here 
determine numerically the nominal values for $t_1$ and $t_2$. For $t_1$, we set $t_{1o}$ as the time when the separation
between the right and left wave packets is maximized. We then set $t_1 = t_{1o} + \delta_1/\omega$. 
For $t_2$, we set $t_{2o}$ as the time
at which the distance between the two interfering wave packets is a minimum, after making $n$ orbits around the trap. 
Specifically, we minimize the quantity
\beq
\delta r^2 = |\vr_{R+}-\vr_{R-}|^2 + |\vr_{L+}-\vr_{L-}|^2
\eeq
where the positions are labeled as in Eq.~\eqref{positions}. We then set $t_2 = t_{2o} + \delta_2/\omega$. This
procedure mimics the way the timing would be determined experimentally. In total, we
include forty-three perturbation parameters in the analysis, all of which are dimensionless.

The classical trajectories themselves are determined using the MATLAB ode45 solver, with tolerance parameters of $10^{-8}$.
The solver is simultaneously used to
integrate the dynamic phase terms of Eq.~\eqref{phidyn}. The initial and final points of the trajectories are used to determine
$\phi^{\text{laser}}$ and $\phi^{\text{sep}}$ via Eqs.~\eqref{laserphase} and \eqref{sepphase}. From these, the total
differential phase $\Phi$ is calculated. We then vary the expansion parameters to determine their sensitivity. We numerically
calculate both the first derivatives $\partial\Phi/\partial\eta_i$ and second derivatives $\partial^2\Phi/\partial\eta_i\partial\eta_j$ for 
expansion parameters $\{\eta_i\}$. We use an increment step $\Delta\eta = 10^{-4}$ and we 
estimate the derivative values to have a numerical accuracy of $10^{-4}$ or better.

\begin{table}
\caption{Sensitivity of the differential phase $\Phi$ to small parameters $\eta_i$. The phase is normalized by $kR$ for 
Bragg laser wave number $k$ and nominal orbit radius $R$. The integer number of orbits is $n$. 
Here we show the thirty significant terms observed for a spherically symmetric trap with $\zeta = 1$. The entries below the horizontal line
in the second column depend on $\zeta$, as seen in Table~\protect\ref{table2}. \label{table1}}
\renewcommand{\arraystretch}{2.3}
\begin{tabular}{ccl||ccl}
\toprule
$ \eta_1$ & $\eta_2$ & $\displaystyle \frac{1}{kR} \frac{\partial^2\Phi}{\partial\eta_1\partial\eta_2}$ & $ \eta_1$ & $\eta_2$ & $\displaystyle \frac{1}{kR} \frac{\partial^2\Phi}{\partial\eta_1\partial\eta_2}$ \\[1mm] \hline
$ \delta_1 $	&	$ c_{110} $	&	$\displaystyle 2\pi n$	&	$ c_{400} $	&	$ c_{310} $	&	$\displaystyle -\frac{9\pi^2}{8} n^2 $	\\
$ \delta_2 $	&	$ c_{110} $	&	$\displaystyle \frac{\pi}{2}(1+4n)$	&	$ c_{400} $	&	$ c_{130} $	&	$\displaystyle -\frac{9\pi^2}{8}n(1+3n)$	\\
$ c_{200} $	&	$ c_{110} $	&	$ \displaystyle -\pi^2 n^2$	&	$ c_{220} $	&	$ c_{310} $	&	$\displaystyle \frac{3\pi^2}{8}n^2 $	\\
$ c_{020} $	&	$ c_{110} $	&	$\displaystyle \pi^2 n^2$	&	$ c_{220} $	&	$ c_{130} $	&	$\displaystyle \frac{3\pi^2}{8} n^2 $	\\
$ \delta_1 $	&	$ c_{310} $	&	$\displaystyle -\frac{3\pi}{2}n $	&	$ c_{040} $	&	$ c_{310} $	&	$\displaystyle -\frac{9\pi^2}{8} n^2 $	\\
$ \delta_1 $	&	$ c_{130} $	&	$\displaystyle \frac{3\pi}{2} n $	&	$ c_{040} $	&	$ c_{130} $	&	$\displaystyle \frac{9\pi^2}{8} n^2 $	\\ \cline{4-6}
$ \delta_2 $	&	$ c_{310} $	&	$\displaystyle \frac{3\pi}{2}n $	&	$ c_{011} $	&	$ c_{101} $	&	$ \displaystyle\frac{\pi^2}{4} n(1+2n)$	\\
$ \delta_2 $	&	$ c_{130} $	&	$\displaystyle \frac{3\pi}{8}(1+4n) $	&	$ c_{011} $	&	$ c_{301} $	&	$\displaystyle -\frac{3\pi^2}{16}n(1+2n) $	\\
$ c_{200} $	&	$ c_{310} $	&	$\displaystyle \frac{3\pi^2}{4} n^2 $	&	$ c_{011} $	&	$ c_{121} $	&	$\displaystyle \frac{\pi^2}{16} n(1+2n) $	\\
$ c_{200} $	&	$ c_{130} $	&	$\displaystyle -\frac{3\pi^2}{4}n^2 $	&	$ c_{101} $	&	$ c_{211} $	&	$\displaystyle \frac{3\pi^2}{8} n^2 $	\\
$ c_{110} $	&	$ c_{400} $	&	$\displaystyle -\frac{3\pi^2}{2} n(1+3n) $	&	$ c_{101} $	&	$ c_{031} $	&	$\displaystyle \frac{3\pi^2}{16}n(1+2n) $	\\
$ c_{110} $	&	$ c_{220} $	&	$\displaystyle \frac{\pi^2}{2} n^2 $	&	$ c_{301} $	&	$ c_{211} $	&	$\displaystyle \frac{3\pi^2}{32} n^2$	\\
$ c_{110} $	&	$ c_{040} $	&	$\displaystyle \frac{3\pi^2}{2} n^2 $	&	$ c_{301} $	&	$ c_{031} $	&	$\displaystyle -\frac{9\pi^2}{64} n(1+2n) $	\\
$ c_{020} $	&	$ c_{310} $	&	$\displaystyle -\frac{3\pi^2}{4} n^2 $	&	$ c_{211} $	&	$ c_{121} $	&	$\displaystyle \frac{3\pi^2}{32} n^2$	\\
$ c_{020} $	&	$ c_{130} $	&	$\displaystyle \frac{3\pi^2}{4}n^2 $	&	$ c_{031} $	&	$ c_{121} $	&	$\displaystyle \frac{3\pi^2}{64}n(1+2n) $	\\

\midrule
\bottomrule
\end{tabular}
\end{table}

\begin{table}
\caption{Phase sensitivity for a cylindrically symmetric trap with $\omega_z = \zeta \omega$.
Terms below the horizontal line in the second column are those that that differ from their values in the spherically symmetric case of Table~\protect\ref{table1}. 
Other terms from Table~\protect\ref{table1} apply without change. Terms in the first column and above the line are those that appear only in the cylindrical case. 
The dependence on $\zeta$
and on the number of orbits $n$ are determined numerically. The $n$-dependence is parametrized by the functions 
$f_1(n) = \sin 2\pi\zeta(n+1/4) - \sin \pi\zeta/2$, $f_2(n) = \cos 2\pi\zeta(n+1/4) - \cos \pi\zeta/2$, and $f_3(n) = 1-\cos(2\pi\zeta n)$.
 \label{table2}}
\renewcommand{\arraystretch}{2.3}
\begin{tabular}{ccc||ccc}
\toprule
$ \eta_1$ & $\eta_2$ & $\displaystyle \frac{1}{kR} \frac{\partial^2\Phi}{\partial\eta_1\partial\eta_2}$ & $ \eta_1$ & $\eta_2$ & $\displaystyle \frac{1}{kR} \frac{\partial^2\Phi}{\partial\eta_1\partial\eta_2}$ \\[1mm] \hline
$ \psi_x'' $	&	$ \psi_y'' $	&	$ \displaystyle -\frac{4f_1(n)}{\zeta} $	&	$ \psi_y'' $	&	$ c_{121} $	&	$\displaystyle \frac{2f_1(n)(3-\zeta^2)}{\zeta(1-\zeta^2)(9-\zeta^2)} $	\\[1mm] \cline{4-6}
$ \psi_x'' $	&	$ c_{011} $	&	$ \displaystyle \frac{2f_1(n)}{\zeta(1-\zeta^2)}  $	&	$ c_{011} $	&	$ c_{101} $	&	$ \displaystyle -\frac{f_1(n)}{\zeta(1-\zeta^2)^2} $	\\
$ \psi_y'' $	&	$ c_{101} $	&	$ \displaystyle \frac{2f_1(n)}{\zeta(1-\zeta^2)} $	&	$ c_{011} $	&	$ c_{301} $	&	$ \displaystyle \frac{6f_1(n)}{\zeta(1-\zeta^2)^2(9-\zeta^2)} $	\\
$ \displaystyle \frac{z_0}{R} $	&	$ c_{111} $	&	$ \displaystyle \frac{2f_2(n)}{4-\zeta^2} $	&	$ c_{011} $	&	$ c_{121} $	&	$\displaystyle -\frac{f_1(n)(2-\zeta^2)}{\zeta(1-\zeta^2)^2(9-\zeta^2)} $	\\
$ \displaystyle \frac{v_{z0}}{v_B} $	&	$ c_{111} $	&	$ \displaystyle \frac{2f_1(n)}{\zeta(4-\zeta^2)} $	&	$ c_{101} $	&	$ c_{211} $	&	$ \displaystyle \frac{6f_3(n)}{(1-\zeta^2)^2(9-\zeta^2)}  $	\\
$ c_{201} $	&	$ c_{111} $	&	$\displaystyle  - \frac{6f_2(n)}{\zeta^2(4-\zeta^2)^2} $	&	$ c_{101} $	&	$ c_{031} $	&	$ \displaystyle -\frac{6f_1(n)}{\zeta(1-\zeta^2)^2(9-\zeta^2)} $	\\
$ c_{021} $	&	$ c_{111} $	&	$\displaystyle \frac{2f_2(n)}{\zeta^2(4-\zeta^2)} $	&	$ c_{301} $	&	$ c_{211} $	&	$\displaystyle \frac{12 f_3(n)}{(1-\zeta^2)^2(9-\zeta^2)^2} $	\\
$ \psi_x'' $	&	$ c_{211} $	&	$\displaystyle  -\frac{12f_2(n)}{(1-\zeta^2)(9-\zeta^2)} $	&	$ c_{301} $	&	$ c_{031} $	&	$ \displaystyle \frac{36f_1(n)}{\zeta(1-\zeta^2)^2(9-\zeta^2)^2} $	\\
$ \psi_x'' $	&	$ c_{031} $	&	$\displaystyle  \frac{12f_1(n)}{\zeta(1-\zeta^2)(9-\zeta^2)} $	&	$ c_{211} $	&	$ c_{121} $	&	$\displaystyle \frac{6f_3(n)(3-\zeta^2)}{(1-\zeta^2)^2(9-\zeta^2)^2} $	\\
$ \psi_y'' $	&	$ c_{301} $	&	$ \displaystyle -\frac{12f_1(n)}{\zeta(1-\zeta^2)(9-\zeta^2)} $	&	$ c_{031} $	&	$ c_{121} $	&	$\displaystyle -\frac{6 f_1(n)(3-\zeta^2)}{\zeta(1-\zeta^2)^2(9-\zeta^2)^2} $	\\

\midrule
\bottomrule
\end{tabular}
\end{table}

\begin{table}
\caption{Third-order sensitivities in the case of a trap with perfect cylindrical symmetry, for $n$ orbits with $\zeta = 1$. 
 \label{table3}}
\renewcommand{\arraystretch}{2.3}
\begin{tabular}{cccc||cccc}
\toprule
$ \eta_1$ & $\eta_2$ &  $\eta_3$ &  $\displaystyle \frac{1}{kR} \frac{\partial^3\Phi}{\partial\eta_1\partial\eta_2\partial\eta_3}$  &
$ \eta_1$ & $\eta_2$ &  $\eta_3$ &  $\displaystyle \frac{1}{kR} \frac{\partial^3\Phi}{\partial\eta_1\partial\eta_2\partial\eta_3}$ 
\\[1mm] \hline
$\psi_x'$	&	$ c_{400}$	&	$ c_{400}$	&	$\displaystyle -\frac{11\pi^2}{4}n^2$	&	$\delta_2$	&	$\delta_2$	&	$\psi_x'$	&	$4$	\\
$\psi_y'$	&	$ c_{400}$	&	$ c_{400}$	&	$\displaystyle -\frac{\pi^2}{4}n(27-11n)$	&	$\delta_2$	&	$\delta_2$	&	$\psi_y'$	&	$-4$	\\
$\delta_1$	&	$\psi_x'$	&	$ c_{400}$	&	$-2\pi n$	&	$\displaystyle \frac{x_0}{R}$	&	$\displaystyle \frac{v_{y0}}{v_B}$	&	$c_{400}$	&	$\pi n$	\\
$\delta_1$	&	$\psi_y'$	&	$ c_{400}$	&	$-11\pi n$	&	$\displaystyle \frac{y_0}{R}$	&	$\displaystyle \frac{v_{x0}}{v_B}$	&	$c_{400}$	&	$\pi n$	\\
$\delta_1$	&	$\delta_2$	&	$\psi_x'$	&	$4$	&	$\delta_2$	&	$\psi_x'$	&	$c_{400}$	&	$\pi n$	\\
$\delta_1$	&	$\delta_2$	&	$\psi_y'$	&	$-4$	&	$\delta_2$	&	$\psi_y'$	&	$c_{400}$	&	$\displaystyle \frac{\pi}{8}(9-8n)$	\\

\midrule
\bottomrule
\end{tabular}
\end{table}

We first consider the case of a spherically symmetric potential with $\zeta = 1$ in Eq.~\eqref{potentialterms}. We observe no first-order
dependence on the expansion parameters, but we observe thirty significant second-order
terms, displayed in Table~\ref{table1}. Although these are numerical results, we find they reduce to simple fractions
when appropriate factors of $\pi$ are included. We determined the $n$ dependence by fitting the results to low-order
polynomials, again obtaining simple integer coefficients.
The terms listed in the table are all those with magnitudes larger than $10^{-4}$ at $n=1$.
The dependence on 
$\delta_1$, $\delta_2$ and $c_{110} = 2\gamma$ agrees with the analytic results from Eq.~\eqref{phi_appx}. 
We observe dependencies on
$(c_{110}, c_{200})$ and $(c_{110},c_{020})$, corresponding to $\gamma$ and $\Delta$ in Eq.~\eqref{harmpot},
which are not present in Eq.~\eqref{phi_appx};
this arises due to the difference in how $\delta_1$ and $\delta_2$ are treated. Of the forty-three parameters considered, we find that sixteen
contribute to second-order terms.

\begin{figure}
	\includegraphics[width=13.5cm]{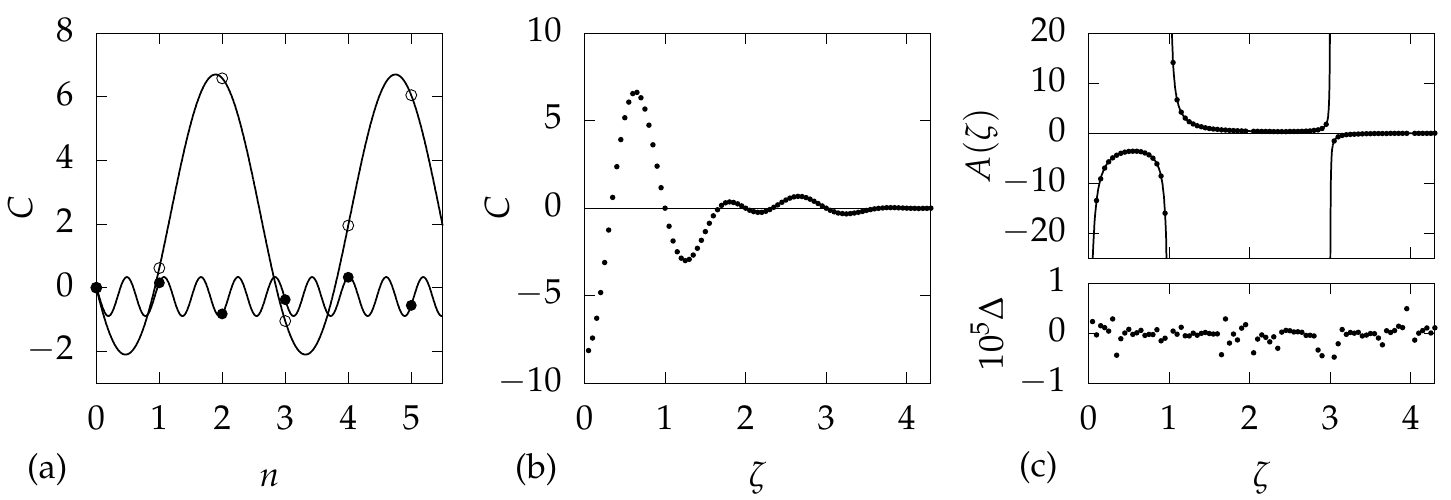}
	\caption{Analysis of numerical results, for the example term $C = (1/kR) \partial^2\Phi/\partial\eta_1\partial\eta_2$ 
	with $\eta_1 = \psi_y''$ and $\eta_2 = c_{301}$. 
	(\textbf{a}) Variation of $C$ with orbit number $n$. The open points show results for $\zeta = 0.35$, and the filled points for $\zeta = 1.7$. 
	From the analytical results, we anticipate that the phase will vary sinusoidally with $2\pi\zeta n$, and we also know that $\Phi = 0$ at $n  = 0$. 
	We therefore fit the points to a function 	$A[\sin(2\pi\zeta(n+n_0))-\sin(2\pi\zeta n_0)]$, with the results shown as curves. While the amplitude $A$
	depends on $\zeta$, we obtain $n_0 = 1/4$ in all cases. 
	(\textbf{b}) Variation of $C$ with $\zeta$, for $n = 1$. This illustrates that the result is a smooth function over the range considered. 
	(\textbf{c}) Points in the upper plot
	show the variation with $\zeta$ of the amplitude $A = C/[\sin(2\pi\zeta(n+1/4))-\sin(\pi\zeta/2)]$. The graph exhibits obvious
	poles at $\zeta = 0$, $\zeta = 1$, and $\zeta = 3$. We incorporate these polls into a rational fit function $A_0/[(\zeta)(1-\zeta^2)(9-\zeta^2)]$
	and find good agreement for constant $A_0 = -12$. The curves show this fit. In the lower plot, $\Delta$ is the fit residual
	$A +12/[(\zeta)(1-\zeta^2)(9-\zeta^2)]$. At larger $\zeta$, the residuals grow but the coefficient $C$ is very small.
		\label{termfig}}
\end{figure}

We also consider the case of a cylindrically symmetric trap, with $\zeta \neq 1$. Most of the terms reported in 
Table~\ref{table1} are unchanged, so Table~\ref{table2} reports only those which are different.
Terms in the first column and above the line in the second column do not appear in Table~\ref{table1}. Terms
below the line are those from from Table~\ref{table1} that are observed to depend on $\zeta$. 
Here we map out the dependence on both $n$ and $\zeta$ by
calculating a range of values and guessing appropriate fitting functions; again we find simple numerical coefficients.
An example of this analysis is described in Fig.~\ref{termfig}. We find that
the term involving $\psi_x''$ and
$\psi_y''$ agrees with the analytic result of \eqref{phi_appx}.
Note that although the terms in the table exhibit poles in $\zeta$,
the divergences are all canceled by zeros of the $f_i(n)$ functions. We confirmed that the limits for $\zeta \rightarrow 1$ agree with the
values in Table~\ref{table1}. As before, the table includes all terms with magnitudes larger than $10^{-4}$ at $n=1$. Here we find 
an additional seven parameters contributing to the phase.   

The total of twenty-three parameters that contribute in second order 
fall into two independent groups, with no phase terms $(\eta_1,\eta_2)$ 
drawing from both groups. The terms above the line in Table~\ref{table1} depend on ten parameters 
$\{\delta_1, \delta_2, c_{200}, c_{020}, c_{110}, c_{400}, c_{310},$
$ c_{220}, c_{130}, c_{040}\}$
that characterize the horizontal motion only. The remaining thirteen parameters
involve coupling to the vertical motion. Of this second group, six contribute
when $\zeta = 1$, but only three, $v_{z0}, c_{111}$ and $c_{201}$, contribute
when $\zeta = 2$. None contribute when $\zeta = 4$ or a larger integer. Reducing the 
number of sensitive parameters is useful since it makes the rotation sensor
more robust.

It is interesting to consider which perturbation terms would be present in a trap that 
maintained perfect cylindrical symmetry. This symmetry requires $c_{200} = c_{020}$,  
$c_{201} = c_{021}$, $c_{202} = c_{022}$, and $c_{400} = c_{040} = c_{220}/2$;
other potential terms involving $x$ and $y$ must be zero. In this case, the only surviving second-order
term is the $(\psi_x'', \psi_y'')$ dependence from Table~\ref{table2},
which can itself be eliminated if $\zeta$ is close to an integer. Using these symmetry constraints to reduce the parameter space, we also explored
the third-order dependence of $\Phi$. Table~\ref{table3} lists the twelve largest terms observed, in 
the case $\zeta = 1$.
Here the largest omitted term has a magnitude (at $n =1$) that is nine times smaller than the smallest term included. From the results, it is
evident that the horizontal quartic anharmonicity $c_{400}$ is particularly important at this order.

\section{Implications for Experiments}

Phase shifts arising from imperfections in the trapping potential will limit 
the accuracy of the interferometer's performance as a rotation sensor. If we interpret
a second-order perturbation from the potential as a rotation error $\delta\Omega$, then we evaluate the corresponding
Sagnac phase as
\beq
\frac{8\pi n R^2 m}{\hbar} \delta\Omega = kR C_{ij} \eta_i \eta_j,
\eeq
where $C_{ij} = (1/kR) \partial^2\Phi/\partial\eta_i\partial\eta_j$ is an entry from Tables~\ref{table1} or \ref{table2}. 
Using $R = 2\hbar k/m\omega$, we have
\beq
\delta \Omega = \frac{C_{ij}}{16\pi n} \omega \eta_i \eta_j.
\eeq
The largest coefficients $C$ have magnitudes that are comparable to $16\pi n^2$, so we conclude
that the effective rotation error is approximately $\delta \Omega \approx n\omega\eta_i\eta_j$. This indicates the level of trap
imperfections that can be tolerated for a given rotation accuracy. For instance, achieving an accuracy $\delta\Omega$
of order $10^{-9}$ rad/s in a trap with $\omega = 2\pi\times 2$~Hz and $n = 1$ would require the $\{\eta_i\}$ to be of order $10^{-5}$.
This result also indicates that the sensitivity to trap imperfections generally increases with $n$ and $\omega$, so it is better to use 
a single orbit in a weaker trap, as opposed to multiple orbits in a tighter trap to achieve the same Sagnac 
area.

It is experimentally feasible to implement a magnetic trap that is stable to a part in $10^5$ or better \cite{Merkel2019,Xu2019}, but 
it would be challenging to design a trap with imperfections that are zero to this level of accuracy. One approach would be
to accept a static phase offset that can be measured and subtracted out to obtain a pure rotation signal,
but the stability tolerance required for a parameter $\eta_i$ grows more stringent the larger its partner parameter $\eta_j$ is.
Alternatively, if all relevant trap parameters can be adjusted experimentally, the interferometer itself can be used
to set the parameters to zero. The second-order phase error is
\beq
\Phi \approx \paren{\veta - \veta_0}^\dagger M \paren{\veta-\veta_0}
\eeq
where here $\veta$ is a vector of experimental parameters and $\veta_0$ contains the  (initially unknown)
parameter values for which the interferometer configuration is ideal.
The matrix $M$ is composed of second derivatives $\partial^2\Phi/\partial\eta_i\partial\eta_j$, which can either be calculated 
as in the previous sections or measured experimentally by varying the parameters in pairs and observing the phase response. 
In a similar way, the gradient vector $\nabla\Phi = \{\partial\Phi/\partial\eta_i\}$ can be measured experimentally at an initial
value of $\veta = 0$. Since $\nabla\Phi \approx 2M(\veta-\veta_0)$, we can obtain an estimate for $\veta_0$ as
\beq
\veta_0 \approx -\frac{1}{2} M^{-1} \nabla\Phi.
\eeq
The parameters can then be set to this $\veta_0$ and the process can be iterated to converge on the desired parameter set where $\nabla\Phi = 0$.
To this end, it is useful that the parameters fall into independent groups, since this means each group can be
optimized independently.

We have experimentally demonstrated the required measurement procedure using the apparatus of Ref.~\cite{Moan2020}. Here we focus on 
two parameters, the timing error $\delta_2$ and the $xy$ potential term $c_{110} = 2\gamma$. 
Experimental control of $\delta_2 = \omega t_2 - 2\pi$ is straightforward via timing. To control $\gamma$,
we make use of a feature of the time-orbiting potential trap. Our trap uses a rotating bias field
with components
\beq
\vec{B} = {B_0 \cos(\Omega_1 t)\cos\paren{\Omega_2 t + \frac{\beta}{2}} \unitv{x} 
+B_0 \cos(\Omega_1 t)\cos\paren{\Omega_2 t - \frac{\beta}{2}} \unitv{y}
+ B_z \sin(\Omega_1 t) \unitv{z}},
\eeq
where $\Omega_1 = 2\pi\times 10$~kHz, $\Omega_2 = \Omega_1/10$, and $\beta$ is an experimentally adjustable phase.
In combination with an oscillating gradient field and gravity, this produces a time-averaged potential \cite{Moan2020,Moan2020b}
\beq
V = \frac{1}{2}m\omega^2 \paren{x^2 + y^2 + \frac{2}{7}\beta xy + \zeta^2 z^2},
\eeq
where the $\beta$ phase provides the desired control of the $xy$ term. Comparing to Eq.~\eqref{phi_appx} , we see $\gamma = \beta/7$.
The experiment used $^{87}$Rb atoms in a trap with $\omega = 2\pi\times 9.26(3)$~Hz. The Bragg wave number was $k = 2\pi/(780.233~\text{nm})$.

\begin{figure}
	\includegraphics[width=13.8cm]{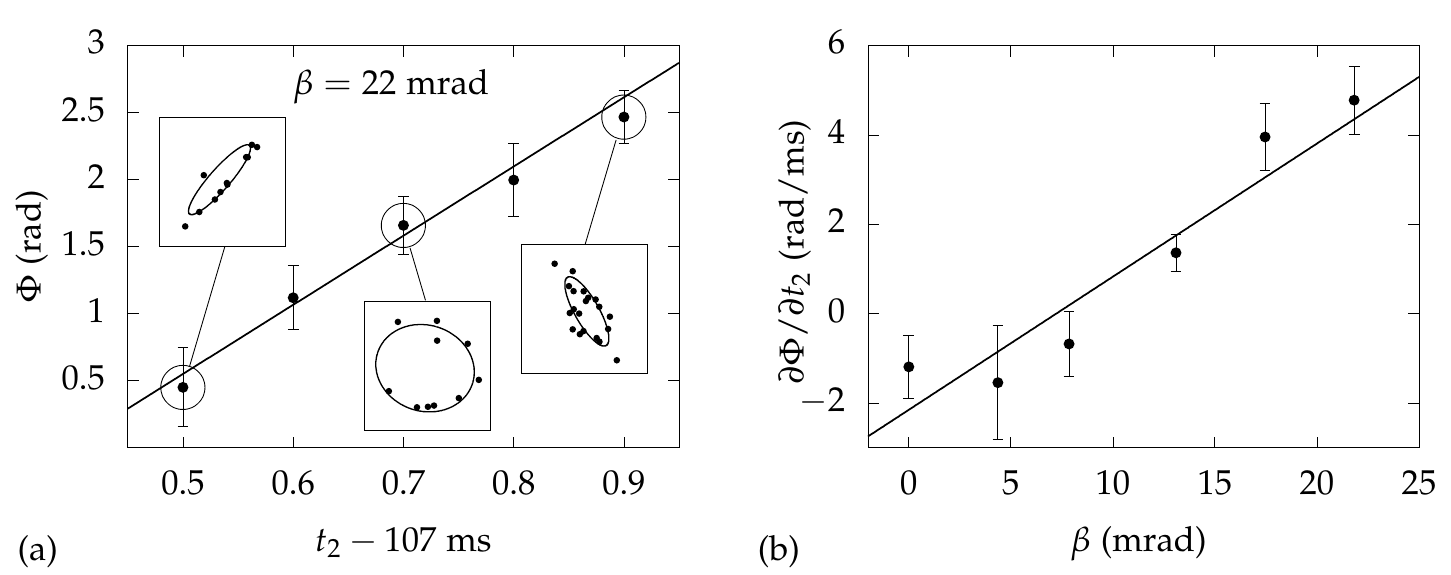}
	\caption{Experimental measurement of phase sensitivity. (\textbf{a}) Variation of differential phase $\Phi$ with interferometer
	time $t_2$. Results are for a trap with magnetic field phase $\beta = 22$~mrad. The line is a linear fit from which the slope
	$\partial\Phi/\partial t_2$ is determined. Insets show plots of $S_L$ vs.\ $S_R$ at the indicated values of $t_2$. The curves 
	are elliptical fits from which the phase values are determined.
	 (\textbf{b}) Slopes from (a) plotted vs.\ $\beta$.
		The line is again a linear fit with slope $\partial^2\Phi/\partial t_2 \partial \beta = 3.0(4)\times 10^5$~s$^{-1}$.
		\label{datafig}}
\end{figure}

The interferometer was operated as in Ref.~\cite{Moan2020}, with no imposed rotation. For set values of $t_2$ and $\beta$, 
the output differential phase $\Phi$ was determined by taking several measurements of the two interferometer
signals $S_R$ and $S_L$, with $S_s = N_{0s}/N_s$ defined as the fraction of atoms returned to rest in interferometer $s$.
As seen in the Fig.~\ref{datafig}(a) insets, the data fall on ellipses when $S_L$ is plotted against $S_R$. 
The location of a point on the ellipse is 
determined by the common mode phase of the two interferometers, which is noisy in our experiment. 
The ellipticity $e$ depends on the
differential phase as $e = [2/(|\sec\Phi|-1)]^{1/2}$, so the phase can be extracted by fitting the data to an ellipse. 

Using this technique, we measured the dependence of $\Phi$ on $t_2$, and found 
a linear variation as seen in Fig.~\ref{datafig}(a). 
We fit these data to determine the slope $\partial\Phi/\partial t_2$,
and repeated the measurements over a range of bias field phases $\beta$. Figure~\ref{datafig}(b) 
shows that $\partial\Phi/\partial t_2$ itself
varies linearly with $\beta$, so from the slope in Fig.~\ref{datafig}(b) we determine $\partial^2\Phi/\partial t_2 \partial \beta = 3.0(4)\times 10^5$~rad/s.
This corresponds to 
\beq
\frac{\partial^2\Phi}{\partial \delta_2 \partial \gamma} = \frac{7}{\omega} \frac{\partial^2\Phi}{\partial t_2 \partial \beta} = 3.6(5)\times 10^4.
\eeq
In comparison, Eq.~\eqref{phi_appx} predicts for $n = 1$ that $\partial^2\Phi/\partial \delta_2 \partial \gamma = 5\pi kv_B/\omega = 2.5\times 10^4$.
The measurement and calculation differ by $2.2\sigma$, which is ambiguous in terms of agreement. We are currently developing a new
apparatus that will substantially improve the measurement precision and allow a more definitive test of the model.

If $\delta_2$ and $\gamma$ were the only trap imperfections to consider, then it would be possible to establish the 
experimental settings where both parameters were zero, as the point where $\partial \Phi/\partial \delta_2 = \partial\Phi/\partial \gamma = 0$. 
However, Table~\ref{table1} indicates that $\delta_2$ and $c_{110}$ are also coupled to $\delta_1$ and the horizontal quartic anharmonicities. 
Since these variables have not been considered, we cannot expect that $\partial\Phi/\partial c_{110} = 0$ when $\delta_2 = 0$ here. 

We do have some information about the trap anharmonicity. As described in Ref.~\cite{Moan2020a}, the potential can be
characterized using the observed packet trajectories. 
Using $R = 0.2$~mm, we find $c_{201} \approx c_{021} = 0.10(3)$, $c_{003} = 0.09(1)$, $c_{400} \approx c_{040} \approx c_{220}/2 = -0.006(1)$,
$c_{202} \approx c_{022} = 0.12(04)$ and $c_{004} = 0.09(3)$. Although these terms are not very small, the numerical model predicts that they do
not significantly change the expected value of $\partial^2\Phi/\partial \delta_2 \partial \gamma$ determined above.

\section{Conclusions}

The methods presented here are generally useful for characterizing the performance of a trapped atom interferometer. The results for our Sagnac interferometer
scheme show which experimental imperfections are most critical, and they provide guidance for how well they must be controlled in order to 
reach a desired level of rotation sensitivity. It is promising that the system is primarily sensitive to parameters that break the cylindrical symmetry,
since these parameters will be naturally small in an experimental design which is nominally symmetric. Nonetheless, the parameters will need to be
be controlled very precisely to reach state-of-the-art performance levels.

It may also be possible to develop operational protocols which help reduce the sensitivity to imperfections. For instance, the interferometer can
instead be operated by splitting first along $x$ and then along $y$. This would alter many of the trap phase terms but not the 
Sagnac phase, so comparing the two results could reduce the sensitivity to perturbations. We hope to explore this and other schemes in future work.

\vspace{6pt}

\noindent\textbf{Author Contributions:} {Conceptualization, C.S.; methodology, Z.L., E.M. and C.S.; software, Z.L. and C.S.; validation, Z.L, E.M and C.S.; formal analysis, Z.L., E.M. and C.S.; investigation, Z.L, E.M and C.S;  writing---original draft preparation, C.S.; writing---review and editing, Z.L. and E.M.; visualization, E.M. and C.S.; supervision, C.S.; project administration, C.S.; funding acquisition, C.S.. All authors have read and agreed to the published version of the manuscript.}

\vspace{6pt} 

\noindent\textbf{Funding:} {This research was funded by the Defense Advanced Research Projects Agency grant number FA9453-19-1-0007, National Science Foundation grant number  PHY-1607571 and NASA grant number RSA1549080.}

\vspace{6pt} 

\noindent\textbf{Acknowledgements:} {We are pleased to acknowledge advice and comments on the manuscript from M. Edwards and J. Stickney, and we acknowledge assistance with the experimental measurements from A. Fallon and S. Berl.}

\vspace{6pt} 

\noindent\textbf{Conflicts of Interest:} {The authors declare no conflict of interest.} 


\end{document}